\documentclass[10pt]{proc}
\usepackage{authblk}
\usepackage{cite}
\usepackage{graphicx}
\usepackage{url}
\usepackage{amsmath}
\usepackage{multirow}
\usepackage{amssymb}
\usepackage{algorithmicx}
\usepackage{algorithm}
\usepackage{inconsolata}
\usepackage{booktabs}

\def\NoNumber#1{{\def\alglinenumber##1{}\State #1}\addtocounter{ALG@line}{-1}}
\algrenewcommand\alglinenumber[1]{\scriptsize #1:}

\author{Martin Junghanns, Andr\'{e} Petermann, Kevin G\'{o}mez and Erhard Rahm}
\affil{University of Leipzig \& ScaDS Dresden/Leipzig}
\date{}

\begin{document}
\title{GRADOOP: Scalable Graph Data Management\\and Analytics with Hadoop\\\begin{large}- Technical Report -\end{large}}
%

\maketitle

\begin{abstract}
Many Big Data applications in business and science require the management and analysis of huge amounts of graph data. Previous approaches for graph analytics such as graph databases and parallel graph processing systems (e.g., Pregel) either lack sufficient scalability or flexibility and expressiveness. We are therefore developing a new end-to-end approach for graph data management and analysis based on the Hadoop ecosystem, called Gradoop (\textbf{Gra}ph analytics on Ha\textbf{doop}). Gradoop is designed around the so-called Extended Property Graph Data Model (EPGM) supporting semantically rich, schema-free graph data within many distinct graphs. A set of high-level operators is provided for analyzing both single graphs and collections of graphs. Based on these operators, we propose a domain-specific language to define analytical workflows. 
The Gradoop graph store is currently utilizing HBase for distributed storage of graph data in Hadoop clusters. An initial version of Gradoop has been used to analyze graph data for business intelligence and social network analysis.
\end{abstract}

\vspace{5mm}

\section{Introduction}
\label{sec:introduction}

Graphs are simple, yet powerful data structures to model and analyze relations between real world data objects. The flexibility of graph data models and the variety of graph algorithms made graph analytics attractive in different domains, e.g., for web information systems, social networks \cite{curtiss2013unicorn}, business intelligence \cite{petermann2014biiig,rudolf2013synopsis,wang2014pagrol}  or in the life sciences \cite{chao2009biology, ling2014gemini}. Entities such as web sites, users or proteins can be modeled as vertices while their connections are represented by edges in a graph. Based on that abstraction, graph algorithms help to rank websites, to detect communities in social graphs, to identify pathways in biological networks, etc.. The graphs in these domains are often very large with millions of vertices and billions of edges  making efficient data management and execution of graph algorithms challenging. For the graph-oriented analysis of heterogeneous data, possibly integrated from different sources, graphs should also be able to adequately represent entities and relationships of different kinds.

Currently, two major kinds of systems focus on the management and analysis of graph data: graph database systems and parallel graph processing systems. 
Graph database systems, such as Neo4j \cite{neo4j} or Sparksee \cite{sparksee,martinez2011dex}, focus on the efficient storing and transactional processing of graph data where multiple users can access a graph in an interactive way. They support expressive data models, such as the property graph model \cite{rodriguez2010propertygraph} or the resource description framework \cite{klyne2006rdf}, which are suitable to represent heterogeneous graph data. Furthermore, graph database systems often provide a declarative graph query language, e.g., Cypher \cite{cypher} or SPARQL \cite{sparql}, with support for graph traversals or pattern matching. However, graph database systems are typically less suited for high-volume data analysis and graph mining \cite{guo2014graphbenchmarking, mccoll2014graphdbperformance, shao2012managingandmining} and often do not support distributed processing on partitioned graphs which limits the maximum graph size to the resources of a single machine. 

\begin{figure}[t]
\centering
\includegraphics[width=0.47\textwidth]{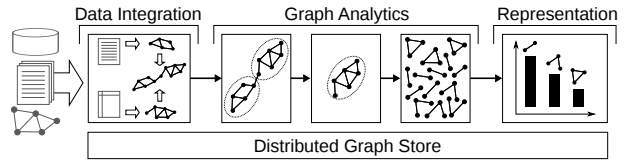}
\caption{Key steps of end-to-end graph analytics}
\label{fig:graph_analytics}
\end{figure}

By contrast, parallel graph processing systems such as Google Pregel \cite{malewicz2010pregel} or Apache Giraph \cite{giraph} process and analyze large-scale graph data in-memory on shared nothing clusters. They provide a tailored computational model where users implement parallel graph algorithms by providing a vertex-centric compute function.
However, there is no support of an expressive graph data model with heterogeneous vertices and edges and high-level graph operators. 
 Parallel in-memory graph processing is also supported by Apache Spark and its GraphX \cite{xin2013graphx} component as well as Apache Flink \cite{alexandrov2014stratosphere}. In contrast to Giraph or Pregel, these systems provide more powerful workflow-based analysis capabilities based on high-level operators for processing and analyzing both graph data as well as other kinds of data. However, these systems also lack support for permanent graph storage and general data management features. Furthermore, there is no support for storing and analyzing many distinct graphs rather than only a single graph.  

The discussion shows that the previous approaches for graph data management and analysis have both strengths and restrictions (more approaches will be discussed in section \ref{sec:related} on related work). We are especially missing support for an end-to-end approach for scalable graph data management and analytics based on an expressive graph data model including powerful analytical operators. Furthermore, we see the need for an advanced graph data model supporting storage and analysis for collections of graphs, 
e.g., for graph comparison in biological applications \cite{chao2009biology} or graph mining in business information networks \cite{petermann2014biiig}. 
The approach should also support the flexible integration of heterogeneous data within a distributed graph store as illustrated in Fig. \ref{fig:graph_analytics}.  

At the German Big Data center of excellence ScaDS Dresden/Leipzig, we have thus 
started with the design and development of the Gradoop (\textbf{Gra}ph Analytics on Ha\textbf{doop}) system for realizing such an end-to-end approach to scalable graph data management and analytics. Its design is based on our previous work on graph-based business intelligence with the so-called BIIIG approach for integrating heterogeneous data within integrated instance graphs \cite{petermann2014biiig, petermann2014demo}. In the ongoing implementation of Gradoop we aim at leveraging existing Hadoop-based systems (e.g., HBase, MapReduce, Giraph) for reliable, scalable and distributed storage and processing of graph data.

In this paper we present the initial design of the Gradoop architecture (section \ref{sec:architecture}) and its underlying data model, the so-called Extended Property Graph Data Model (EPGM, section \ref{sec:datamodel}).  We also outline the implementation of the HBase-based graph data store (section \ref{sec:storage}) and demonstrate the usefulness of our approach for two use cases (section \ref{sec:usecases}). Our main contributions can be summarized as follows:

\begin{itemize}
\item We present the high-level system design of Gradoop, a new Hadoop-based framework for distributed graph data management and analytics. Gradoop aims at an end-to-end approach according to Fig. \ref{fig:graph_analytics} with workflow-based integration of source data into a common graph store, workflow-based graph analysis as well as the visualization of analysis results. 

\item We propose the powerful yet simple EPGM graph data model. It supports both graphs and collections of graphs with heterogeneous vertices and edges  as well as declarative operators for graph analytics. We also show how the operators can be used for the declaration of analytical workflows with a new domain-specific language called GrALa (Graph Analytical Language).    

\item We describe the design and implementation of our distributed graph store built upon Apache HBase. It supports partitioning, replication and versioning of large, heterogeneous graphs.

\item We show the applicability of an initial implementation of Gradoop using two use cases for social network analysis and business intelligence.
\end{itemize}

\begin{figure}
\centering
\includegraphics[width=0.47\textwidth]{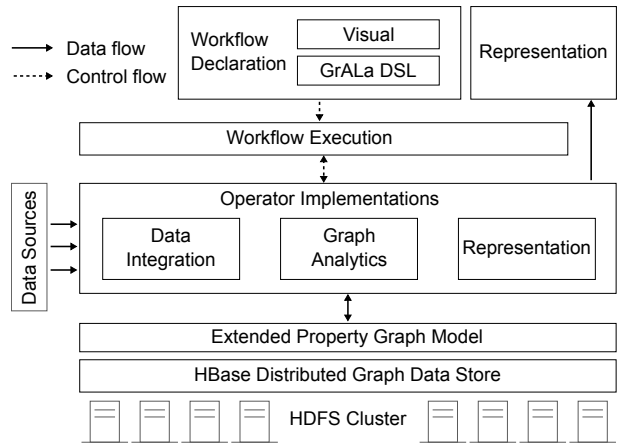}
\caption{Gradoop High-Level Architecture}
\label{fig:architecture}
\end{figure}

\section{Gradoop Architecture}
\label{sec:architecture}

With Gradoop we aim at providing a framework for scalable graph data management and analytics on large, semantically expressive graphs. To achieve horizontal scalability of storage and processing capacity, Gradoop runs on shared nothing clusters and utilizes existing Hadoop-based software for distributed data storage and processing. 

Fig. \ref{fig:architecture} shows the high-level architecture of Gradoop.
Gradoop users can declare data integration and analytical workflows with either a visual interface or by using GrALa, our domain specific language (DSL). Workflows are processed by the workflow execution layer which has access to the actual operator implementations, which in turn access the distributed graph store using the EPGM graph data model. After processing a workflow, results are presented to the user. In the following, we briefly explain the core components and discuss them in more detail in subsequent sections.

\paragraph*{\textbf{Distributed Graph Store}}
The distributed graph store manages a persistent graph database structured according to the EPGM graph data model. It offers basic methods to store, read and modify graph data, and serves as the data source for operators directly operating on the permanent graph representation, e.g., when using MapReduce. Furthermore, the graph output of operators can be written to the data store. 

Efficient graph processing demands a fast retrieval of vertices and their neighborhood as well as a low-latency communication between connected vertices. The graph data has to be physically partitioned across all cluster nodes  so that data is equally distributed for load balancing. Furthermore, the partitioning should support graph processing with little communication overhead (locality of access). 
The Gradoop graph store also manages different versions of the underlying graph, e.g., to  enable the analysis of graph changes over time. Finally, the graph store needs to be resilient against failures and avoid data loss. 

Our current storage implementation builds on HBase \cite{hbase}, an open-source implementation of Google's BigTable \cite{chang2008bigtable}, running on the Apache HDFS  (Hadoop Distributed File System) \cite{hadoop}. HBase supports high data availability through redundant data storage and provides data versioning as well as partitioning. Many Hadoop processing components (e.g., MapReduce, Giraph, Flink)  have built-in support for HBase thererby simplifying their use for realizing Gradoop functionality. 
In section \ref{sec:storage} we describe the Gradoop graph store in more detail. 

\paragraph*{\textbf{Extended Property Graph Data Model}}
Our extended property graph data model (EPGM) describes how graph databases are structured and defines a set of declarative operators. EPGM is an extension of the property graph model \cite{rodriguez2010propertygraph}, which is used in various graph database systems. To facilitate integration of heterogeneous data from different sources, it does not enforce any kind of schema, but the graph elements (vertices, edges, logical graphs) can have different type labels and attributes. For enhanced analytical expressiveness, EPGM supports multiple logical graphs inside a graph database. Logical graphs are, as well as vertices and edges, first-class citizens of our data model which can have their own properties. Furthermore, collections of logical graphs can be the input and output of analytical operators. The details of our data model and its analytical operators are discussed in section \ref{sec:datamodel}.

\paragraph*{\textbf{Operator Implementations}}
The EPGM operators need to be efficiently implemented for use in analytical workflows. This also holds for further operators for presenting analysis results and to perform data import, transformation and integration tasks to load external data  into the Gradoop graph store. All these operators have access to the common graph store and need to be executed in parallel on the underlying Hadoop cluster. For the operator implementation Gradoop utilizes existing systems such as MapReduce, Giraph, Flink or Spark thereby taking advantage of their respective strengths. For example, MapReduce is well suited for ETL-like data transformation and integration tasks while Giraph can efficiently process graph mining algorithms. We implemented a first set of operators as a proof of concept and use them in our evaluation in section \ref{sec:usecases}.

\paragraph*{\textbf{Workflow Execution}}
The workflow execution component is responsible for managing the complete execution of data integration workflows or analytical workflows. Before a declared workflow can be executed, it is transformed into an executable program. The workflow execution has access to the operator implementations and runs and monitors their execution. Furthermore, it manages intermediate operator results and provides status updates to the user. At the beginning of a workflow, as necessary for the first operator, the graph or parts of it are read from the graph store by the execution system. Intermediate results are either written to the graph store or are cached in memory by the execution layer. The latter case will be preferred for high performance and especially used if two subsequent operators are implemented in the same system, e.g., Spark or Flink. Final analysis results are stored or forwarded to the presentation layer.
\pagebreak
\paragraph*{\textbf{Workflow Declaration and Result Representation}}
The typical Gradoop users are data scientists and analysts. They are responsible for the specification of data integration and analysis workflows and evaluate the obtained results.
For ease of use, workflows can be declaratively specified by writing a DSL (GrALa) script using the available operators which is then handed over to the workflow execution subsystem. Alternatively, workflows are visually defined using a browser-based front-end and then automatically transformed into a DSL script. Workflow results are either represented through graph visualization (e.g., colored subgraphs or specific layouts) or combined with charts and tables (e.g., aggregate values for subgraphs or frequency distributions of graph patterns).

As mentioned in the introduction, the implementation of Gradoop is still going on, so that some of the introduced components, in particular the workflow execution, the visual workflow definition and result representation still need to be completed. The current focus is on the graph data model, the definition of analytical workflows and the underlying graph data store. For data integration, we will port the approaches proposed for BIIIG \cite{petermann2014biiig, petermann2014demo} to Hadoop and adapt our MapReduce-based Dedoop tool for entity resolution \cite{kolb2012dedoop} to support graph data. 

\begin{figure*}[t]
\centering
\includegraphics[width=\textwidth]{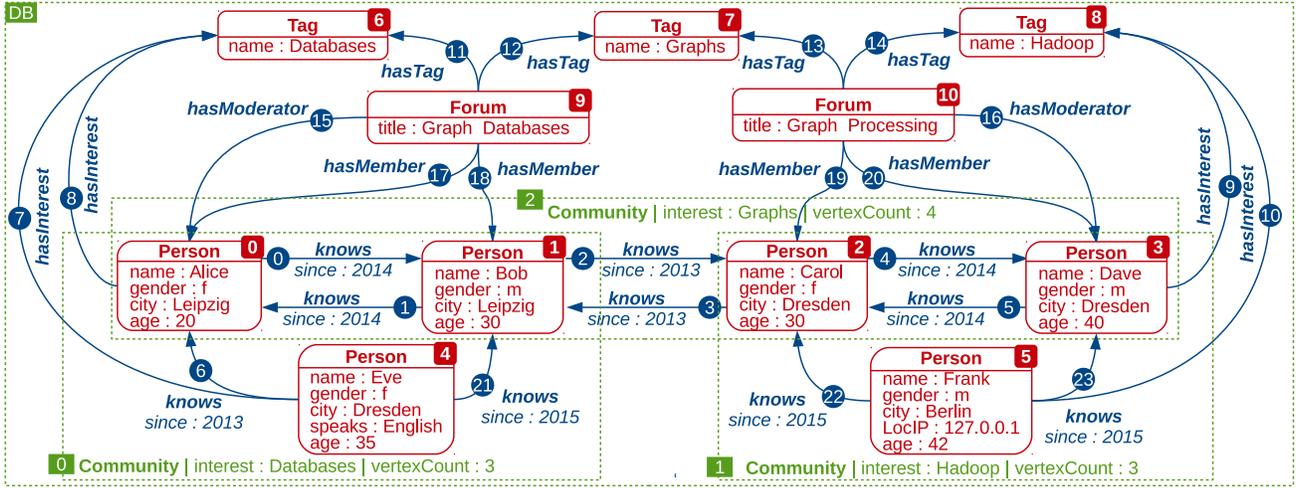}
\caption{Example EPGM database graph}
\label{fig:epg-example}
\end{figure*}

\section{Graph Data Model}
\label{sec:datamodel}
In this section, we introduce the EPGM data model of Gradoop. We first describe the representation of graph data with EPGM and then present its analytical operators.

\subsection{Graph Representation}

The design of EPGM data representation is based on the following requirements that we have derived from various analytical scenarios:

\paragraph*{\textbf{Simple but powerful}}
The graph model should be powerful enough to support the graph structures of most use cases for graph analytics. On the other hand, it should also be intuitive and easy to use. For this reason we favor a model with a flat structure of vertices and binary edges.

\paragraph*{\textbf{Logical graphs}}
Support for more than one graph in the data model is advantageous since many analytical applications involve multiple graphs, such as communities in social networks or multiple executions of a business process. These graphs may have common vertices, edges or subgraphs.

\paragraph*{\textbf{Type labels and attributes}}
Graph data from real-world scenarios is often heterogeneous exhibiting multiple types of vertices, edges and graphs. A graph model should thus support different types and heterogeneous attributes for all of its elements in a uniform way. Additionally, the meaning of relationships requires edges to be directed.

\paragraph*{\textbf{Loops and parallel edges}}
In many real-world scenarios there may be self-connecting edges or multiple edges having the same incident vertices, for example, to describe different relationships between persons. Hence, a graph data model should support loops and parallel edges.

\vspace{5mm}

In its simplest form, a graph  $G = \langle V,E \rangle$  consists of a set of vertices $V$ and a set of binary edges $E \subseteq V \times V$.  Several extensions of this simple graph abstraction have been proposed to define a graph data model  \cite{angles2012comparison,angles2008survey}. One of those models, the \textit{property graph model (PGM)} \cite{rodriguez2010propertygraph, rodriguez2010traversalpattern}, already meets our requirements in large parts. The PGM is widely accepted and used in graph database systems (e.g., Neo4j), industrial research projects (e.g., SAP Active Information Store \cite{rudolf2013graph}) and in parallel processing systems such as Spark GraphX. A property graph is a directed multigraph supporting encapsulated \textit{properties} (named attributes) for both vertices and edges. Properties have the form of key-value pairs (e.g., \texttt{name:Alice} or \texttt{weight:42}) and are defined at the instance level without requiring an upfront schema definition. However, the PGM foresees type labels only for edges (e.g., \texttt{knows}). Hence, it has no support for multiple graphs and respective graph type labels and graph properties. Furthermore, there are no operators on multiple graphs.  

To meet all of the posed requirements, we have developed the \textbf{E}xtended \textbf{P}roperty \textbf{G}raph \textbf{M}odel (EPGM). In this model, a database consists of multiple property graphs which we call \emph{logical graphs}. These graphs are application-specific subsets from shared sets of vertices and edges, i.e., may have common vertices and edges. Additionally, not only vertices and edges but also logical graphs have a type label and can have different properties. Formally, we define an EPGM database as 

$$
	DB_{EPGM} = \langle \mathcal{V}, \mathcal{E}, \mathcal{G}, T, \tau, K, A, \kappa \rangle.
$$

A (graph) database $DB_{EPGM}$ consists of a vertex space $\mathcal{V} = \langle v_i \rangle$, an edge space $\mathcal{E} = \langle e_k \rangle$ and a set of logical graphs $\mathcal{G} = \langle G_m\rangle$. Vertices, edges and (logical) graphs are identified by the respective indices $i,k,m \in \mathbb{N}$.  An edge $e_k = \langle v_i,v_j \rangle$ with $v_i,v_j \in V$ directs from $v_i$ to $v_j$ and supports loops (i.e., $i = j$). There can be multiple edges between two vertices which are differentiated by distinct identifiers.  A logical graph $G_m = \langle V,E \rangle$ is an ordered pair  of a subset of vertices $V \subseteq \mathcal{V}$ and a subset of edges $E \subseteq \mathcal{E}$. We use $G_{DB}$ to denote the graph of all vertices $V(G_{DB}) = \mathcal{V}$ and all edges $E(G_{DB}) = \mathcal{E}$ of a database. Graphs may potentially overlap such that $\forall G_i,G_j \in \mathcal{G} : \left| V(G_i) \cap V(G_j) \right| \geq 0 \wedge \left| E(G_i) \cap E(G_j) \right| \geq 0$. For the definition of type labels we use label alphabet $T$ and a mapping $\tau : (\mathcal{V} \cup \mathcal{E} \cup \mathcal{G}) \rightarrow T$. Similarly, properties (key-value pairs) are defined by key set $K$, value set $A$ and mapping $\kappa: (\mathcal{V} \cup \mathcal{E} \cup \mathcal{G}) \times K \rightarrow A$.

Figure \ref{fig:epg-example} shows an example EPGM database graph $G_{DB}$ for a simple social network. Formally, $G_{DB}$ contains of the vertex space $\mathcal{V} = \{ v_0, .., v_{10} \}$ and the edge space $\mathcal{E} = \{ e_0,.., e_{23} \}$. Vertices represent persons, forums and interest tags, represented by corresponding type labels (e.g., \texttt{Person}) and further described by their properties (e.g., \texttt{type:Tag} or \texttt{name:Alice}). Edges describe the relationships between vertices and also have type labels (e.g., \texttt{knows}) and properties (e.g., \texttt{since:2013}). The key set $K$ contains all property keys, for example, \texttt{name}, \texttt{title} and \texttt{since}, while the value set $A$ contains all property values, for example, \texttt{"Alice"}, \texttt{"Graph Databases"} and \texttt{2015}. Vertices with the same type label may have different property keys, e.g., $v_4$ and $v_5$.

Furthermore, the sample database contains the logical graph set $\mathcal{G} = \{ G_0, G_1, G_2 \}$, where each logical graph represents a community inside the social network, in this case, specific interest groups (e.g., Graph Databases). Such groups can be found by application-specific subgraph detection algorithms. In this example, users will be part of a community, if they are a member of a forum that is tagged by a specific topic or have direct interest in that topic. Each logical graph has a specific subset of vertices and edges, for example, $V(G_0) = \{ v_0, v_1, v_4 \}$ and $E(G_0) = \{e_0, e_3, e_6, e_{21}\}$. Considering $G_0$ and $G_1$, it can be seen that vertex sets may overlap such that $V(G_0) \cap V(G_1) = \{v_0, v_1\}$. Additionally, also graphs have type labels (e.g., \texttt{Community}) and may have properties, which can be used to describe the graph by annotating it with specific metrics (e.g., \texttt{vertexCount:3}) or general information about that graph (e.g., \texttt{interest:Databases}). Usually, logical graphs are the result of an operator executed in an analytical workflow. If they need to be re-used, logical graphs can be persisted in the graph store.
 
\subsection{Operators}
\label{sec:operators}
The EPGM provides operators for both single graphs as well as collections of graphs; operators may also return single graphs or graph  collections. In the following, we use \textit{collection} and \textit{graph collection} correspondingly. In Gradoop, collections are ordered to support application-specific sorting of collections and position-based selection of graphs from a collection.
Table \ref{tab:operators} lists our analytical operators together with the definitions of their input and output. The table also shows the corresponding syntax for calling the operators in our domain specific language GrALa (\textbf{Gr}aph \textbf{A}nalytical \textbf{La}nguage). Inspired by modern programming languages, we use the concept of higher-order functions in GrALa for several operators (e.g., to use an aggregate or a predicate function as an operator argument). 
 Based on the input of operators, we distinguish between \textit{collection operators} (shown in the top part of Table \ref{tab:operators})
and  \textit{graph operators} as well as between  \textit{unary}  and \textit{binary operators} (single graph/collection vs. two graphs/collections input). There are also some \textit{auxiliary operators} to apply graph operators on collections or to call specific graph algorithms. 
In addition to the listed operators, GrALa provides basic operators to create, read, update and delete graphs, vertices and edges as well as their properties. Since we can store older versions of graphs in the Gradoop graph store, we can read different versions of graphs and their elements, e.g., for time-based analytics. In the following, we discuss the Gradoop operators in more detail.

\paragraph*{\textbf{Collection Operators}}
 Collection operators can be applied on collections of graphs, vertices and edges. In the following, we focus on graphs  as their usage for vertices and edges is analogous.

The \textit{selection} operator $\sigma_\varphi : \mathcal{G}^n \rightarrow \mathcal{G}^n$ for collections selects the graphs from the input collection that meet a user-defined predicate function $\varphi : \mathcal{G} \rightarrow \{true,false\}$. The output is a collection with all qualifying graphs. 
Algorithm \ref{alg:select} shows two examples using the selection operator in GrALa. 
We first define the input collection (line 1) of three logical graphs identified by their unique id (e.g., \texttt{db.G[0]} corresponds to $G_0$) and assign it to the variable \texttt{collection}. The \texttt{db} object is a reference to the database graph $G_{DB}$. The first user-defined predicate function (line 2) will evaluate to true, if the input graph \texttt{g} has a value greater than 3 for property key \texttt{vertexCount}, i.e., we want to find all graphs with more than three vertices. In line 3, we call the select operator to apply the predicate function on the predefined \texttt{collection}. For our example graph in Figure \ref{fig:epg-example}, the result collection only contains \texttt{db.G[2]}. 

\begin{algorithm}[h]
\caption{Selection Example}
\small
\begin{ttfamily}
\begin{algorithmic}[1]
\State collection = <db.G[0],db.G[1],db.G[2]>
\State predicate1 = (Graph g => g["vertexCount"] > 3)
\State result1 = collection.\textbf{select}(predicate1)
\State predicate2 = (Graph g => g["vertexCount"] == 
\NoNumber \ \ g.V.\textbf{select}(Vertex v => 
\NoNumber \ \ \ \ v["age"] > 20).count()))
\State result2 = collection.\textbf{select}(predicate2)
\end{algorithmic}  
\end{ttfamily}
\label{alg:select}
\end{algorithm}

The second example shows that predicates are not limited to graph properties but can be specified by complex functions on graph vertices or edges. 
The predicate function defined in line 4 allows us to select all graphs where all vertices have a property \texttt{age} with a value greater than 20. To achieve that, we access the vertices of the particular graph (i.e., \texttt{g.V}) and apply a further predicate which uses the selection operator on the vertices. The vertex selection determines all vertices satisfying the age condition. The predicate then evaluates to true if the number (i.e., \texttt{count()}) of the resulting vertices equals the total number of vertices stored in property \texttt{vertexCount}. The \texttt{count} function is a predefined aggregate function and will be discussed below. For our example graph, the resulting collection in line 5 only contains \texttt{db.G[1]}.

As shown in Table \ref{tab:operators}, we also support the set-theoretical operators \textit{union}, \textit{intersection} and \textit{difference} on collections. For example, the call \texttt{<db.G[0],db.G[1]>.\textbf{intersect}(<db.G[1],db.G[2]>)} results in a collection \texttt{<db.G[1]>}.

\begin{algorithm}[h]
\caption{Sort and Top Examples}
\small
\begin{ttfamily}
\begin{algorithmic}[1]
\State sortedColl = db.G.\textbf{sortBy}("vertexCount",:desc)
\State topGraphs = sortedColl.\textbf{top}(2)
\end{algorithmic}  
\end{ttfamily}
\label{alg:sorttop}
\end{algorithm}

Furthermore there are operators for eliminating duplicate graphs in collections based on their index (\textit{distinct} operator), for sorting collections (\textit{sort}) and for selecting the first $n$ ($n \in \mathbb{N}$) graphs in a collection (\textit{top}). 
The sort operator returns a collection sorted by a graph property $k$ in either ascending or descending order, denoted by $o$.  Algorithm \ref{alg:sorttop} shows the usage of sort and top in GrALa. 

\begin{table*}[t]
\centering
\footnotesize
\begin{tabular}{lrcrcll} 
\textbf{Operator} & \multicolumn{5}{c}{\textbf{Definition}} &\textbf{GrALa}\\
Selection 	& $\sigma_\varphi$	&:&$\mathcal{G}^n$  & $\rightarrow$ & $\mathcal{G}^n$&\texttt{collection.select(predicateFunction) : Collection}\\
Distinct 	  & $\delta$      	  &:&$\mathcal{G}^n$  & $\rightarrow$ & $\mathcal{G}^n$&\texttt{collection.distinct() : Collection}\\
Sort by     & $\xi_{k,o}$	      &:&$\mathcal{G}^n$ & $\rightarrow$ & $\mathcal{G}^n$&\texttt{collection.sortBy(key,[:asc|:desc]) : Collection}\\
Top 	      & $\beta_n$      	    &:&$\mathcal{G}^n$  & $\rightarrow$ & $\mathcal{G}^n$&\texttt{collection.top(limit) : Collection}\\
Union		& $\cup$                &:&$(\mathcal{G}^n)^2$  & $\rightarrow$ & $\mathcal{G}^n$&\texttt{collection.union(otherCollection) : Collection}\\
Intersection & $\cap$           &:&$(\mathcal{G}^n)^2$  & $\rightarrow$ & $\mathcal{G}^n$&\texttt{collection.intersect(otherCollection) : Collection}\\
Difference 	& $\setminus$      	&:&$(\mathcal{G}^n)^2$  & $\rightarrow$ & $\mathcal{G}^n$&\texttt{collection.difference(otherCollection) : Collection}\\
\hline
Combination   & $\sqcup$                &:&$\mathcal{G}^2$ & $\rightarrow$ & $\mathcal{G}$&\texttt{graph.combine(otherGraph) : Graph}\\
Overlap       & $\sqcap$                &:&$\mathcal{G}^2$ & $\rightarrow$ & $\mathcal{G}$&\texttt{graph.overlap(otherGraph) : Graph}\\
Exclusion     & $-$                     &:&$\mathcal{G}^2$ & $\rightarrow$ & $\mathcal{G}$&\texttt{graph.exclude(otherGraph) : Graph}\\
\hline
Pattern Matching     & $\mu_{G^*, \varphi}$    &:&$\mathcal{G}$ & $\rightarrow$ & $\mathcal{G}^n$ &\texttt{graph.match(patternGraph,predicateFunction) : Collection}\\
\hline
Aggregation   & $\gamma_{k,\alpha}$              &:&$\mathcal{G}$ & $\rightarrow$ & $\mathcal{G}$ &\texttt{graph.aggregate(propertyKey,aggregateFunction) : Graph}\\  
Projection    & $\pi_{\nu, \epsilon}$   &:&$\mathcal{G}$ & $\rightarrow$ & $\mathcal{G}$ &\texttt{graph.project(vertexFunction,edgeFunction) : Graph}\\
Summarization & $\zeta_{g_v, g_e,\gamma_v,\gamma_e}$ &:&$\mathcal{G}$ & $\rightarrow$ & $\mathcal{G}$ &\texttt{graph.summarize(vertexGroupingKeys,vertexAggregateFunction,}\\
              &                         & &              &               &		         &\texttt{\ \ edgeGroupingKeys,edgeAggregateFunction) : Graph}\\
\hline
Apply         & $\lambda_o$	  &:&$\mathcal{G}^n$  & $\rightarrow$ & $\mathcal{G}^n$   &\texttt{collection.apply(unaryGraphOperator) : Collection} \\
Reduce	      & $\rho_o$	    &:&$\mathcal{G}^n$  & $\rightarrow$ & $\mathcal{G}$     &\texttt{collection.reduce(binaryGraphOperator) : Graph} \\                 
Call          & $\eta_{a,P}$	&:& $\mathcal{G} \cup \mathcal{G}^n$ & $\rightarrow$ & $\mathcal{G} \cup \mathcal{G}^n$   &\texttt{[graph|collection].callForGraph(algorithm,parameters) : Graph}  \\
& & & & & & \texttt{[graph|collection].callForCollection(algorithm,parameters):Collection}  \\
%
\end{tabular}
\caption{Overview of analytical operators in Gradoop}
\label{tab:operators}
\end{table*}

\paragraph*{\textbf{Binary Graph Operators}}
We also support set-theoretical operators to determine the union (\textit{combination} operator), intersection (\textit{overlap}) and difference (\textit{exclusion}) of two graphs resulting in a new graph.  For example, the combination operators is useful  to merge previously selected subgraphs into a new graph.  The combination of input graphs $G_1,G_2$ is a  graph $G'$ consisting of the vertex set $V(G') = V(G_1) \cup V(G_2)$ and the edge set $E(G') = E(G_1) \cup E(G_2)$. For our example graph in Figure \ref{fig:epg-example}, the call \texttt{db.G[0].\textbf{combine}(db.G[2])} results in the new graph $G' = \langle \{ v_0, v_1, v_2, v_3, v_4 \}, \{ e_0, .., e_6, e_{21} \} \rangle$.

Similarly, the overlap of graphs $G_1,G_2$ is a  graph $G'$ with vertex set $V(G') = V(G_1) \cap V(G_2)$ and  edge set $E(G') = E(G_1) \cap E(G_2)$. Applying the exclusion operator to  $G_1$ and $G_2$ determines all $G_1$ elements that do not occur in $G_2$, i.e., $V(G') = V(G_1) \setminus V(G_2)$ and $E(G') = \{ (u,v) \in E(G_1) \mid u \in V(G_1) \setminus V(G_2) \wedge v \in V(G_1) \setminus V(G_2) \}$. 
In the example, the call \texttt{db.G[0].\textbf{overlap}(db.G[2])} returns the graph $G' = \langle \{ v_0, v_1 \}, \{e_0, e_1\} \rangle$ while the call \texttt{db.G[0].\textbf{exclude}(db.G[2])} results in $G' = \langle \{ v_4 \}, \emptyset \rangle$.

\pagebreak
\paragraph*{\textbf{Pattern Matching}}
A fundamental operation in graph analytics is the retrieval of subgraphs matching a user-defined pattern, also referred as \textit{pattern matching}. For example, given a social network scenario, an analyst may be interested in all pairs of users that are member of the same forum. We provide the \textit{pattern matching} operator $\mu_{G^*, \varphi} : \mathcal{G} \rightarrow 	\mathcal{G}^n$, where the search pattern consists of a pattern graph $G^*$ and a predicate $\varphi : \mathcal{G} \rightarrow \{true, false\}$. The operator takes a graph $G$ as input and returns a graph collection $\mathcal{G}' = \{G' \subseteq G \mid G' \simeq G^* \wedge \varphi(G') = true \}$ containing all found matches. Generally speaking, the operator finds all subgraphs of the input graph that are isomorphic to the pattern graph and fulfill the predicate. 

\begin{figure}[h]
\centering
\includegraphics[width=0.47\textwidth]{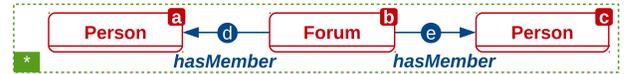}
\caption{Example pattern graph and predicate}
\label{fig:oex-pattern}
\end{figure}

Algorithm \ref{alg:match} shows an example use of our pattern matching operator; the pattern graph is illustrated in Figure \ref{fig:oex-pattern}. For GrALa, we adopted the basic concept of describing graph patterns using ASCII characters from Neo4j Cypher \cite{cypher}, where \texttt{(a)-e->(b)} denotes an edge \texttt{e} that points from a vertex \texttt{a} to a vertex \texttt{b}. In line 1, we describe a \texttt{pattern} of three vertices and two edges, which then can be accessed by variables in isomorphic instances to declare the predicate (e.g., \texttt{graph.V[\$a]}). Property values are accessed using the property key (e.g., \texttt{v["name"]}) or in case of the type label using the reserved symbol \texttt{:type}. In line 2, the \texttt{predicate} is defined as a function which maps a graph to a boolean value. In this function, vertices and edges are accessed by vertex and edge variables and multiple expressions are combined by logical operators. In our example, we compare vertex and edge types to constants (e.g., \texttt{g.V[\$a][:type] == "Forum"}). In line 3, the match operator is called for the database graph \texttt{db} of Figure \ref{fig:epg-example} using \texttt{pattern} and \texttt{predicate} as arguments. 
For the example, the result collection has two subgraphs: $\mathcal{G}' = \{ \langle \{ v_0, v_1, v_9 \}, \{ e_{17}, e_{18} \} \rangle, \langle \{ v_2, v_3, v_{10} \}, \{ e_{19}, e_{20} \} \rangle \}$.

\begin{algorithm}[h]
\caption{Pattern Matching Example}
\small
\begin{ttfamily}
\begin{algorithmic}[1]
\State pattern = new Graph("(a)<-d-(b)-e->(c)")
\State predicate = (Graph g => 
\NoNumber \ g.V[\$b][:type]=="Forum" \&\&
\NoNumber \ g.E[\$d][:type]=="hasMember"\&\&
\NoNumber \ g.V[\$a][:type]=="Person" \&\&
\NoNumber \ g.E[\$e][:type]=="hasMember"\&\& 
\NoNumber \ g.V[\$c][:type]=="Person")
\State result = db.\textbf{match}(pattern,predicate)
\end{algorithmic}  
\end{ttfamily}
\label{alg:match}
\end{algorithm}

\paragraph*{\textbf{Aggregation}} 
An operator often used in analytical applications is aggregation, where a set of values is summarized to a single value of significant meaning. In the EPGM, we support \textit{aggregation} at the graph level by providing the operator $\gamma_{k,\alpha}:\mathcal{G} \rightarrow \mathcal{G}$. Formally, the operator maps an input graph $G$ to an output graph $G'$ and applies the user-defined aggregation function $\alpha : \mathcal{G} \rightarrow \mathbb{R}$. Thus, the resulting graph is a modified version of the input graph with an additional property $k$, such that $\kappa(G',k) \mapsto \alpha(G)$. The resulting property value depends on the applied aggregation function. Basic aggregation functions such as \textit{count}, \textit{sum} and \textit{average} are predefined in GrALa and can be applied on graph, vertex and edge collections (count) and their properties (sum, average), for example, to calculate the average age per community in a social network or the financial result of a business process instance. 

Algorithm \ref{alg:aggregate} shows a simple vertex count example  where the computed cardinality of the vertex set becomes the value of a new graph property \texttt{vertexCount}.

\begin{algorithm}[h]
\caption{Aggregation Example}
\small
\begin{ttfamily}
\begin{algorithmic}[1]
\State g.\textbf{aggregate}(
\NoNumber \ \ "vertexCount",(Graph g => g.V.count()))
\end{algorithmic}  
\end{ttfamily}
\label{alg:aggregate}
\end{algorithm}

\paragraph*{\textbf{Projection}}
The projection operator simplifies a graph representation by keeping only vertex and edge properties necessary for further processing. Furthermore, it is possible to modify (e.g., rename) properties of interest.  For this purpose, the \textit{projection} operator $\pi_{\nu, \epsilon} : \mathcal{G} \rightarrow \mathcal{G}$  applies the bijective projection functions $\nu : \mathcal{V} \rightarrow \mathcal{V}$ and $\epsilon : \mathcal{E} \rightarrow \mathcal{E}$ to an input graph $G$, and outputs the graph $G'$ where $V(G') = \{\nu(v) \mid v \in V(G)\}$, $E(G') = \{\epsilon(e) \mid e \in E(G)\}$ and $G \simeq G'$ (i.e., the input and output graphs are isomorphic). The user-defined projection functions are able to modify type labels as well as property keys and values of vertices and edges, but not their structure. All properties not specified in the projection functions are removed. 

\begin{figure}[h]
\centering
\includegraphics[width=0.47\textwidth]{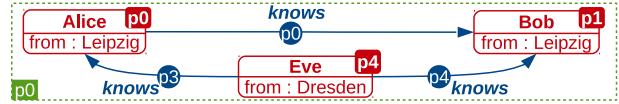}
\caption{Example projection of $G_0$ from figure \ref{fig:epg-example}}
\label{fig:oex-projection}
\end{figure}

Algorithm \ref{alg:project} shows an example GrALa script to project the community graph $G_0$ in Figure \ref{fig:epg-example} to a simplified version shown in Figure \ref{fig:oex-projection}. The vertex function in line 1 determines that all vertex properties are removed, except vertex property \texttt{"city"}, which is renamed to \texttt{"from"}. Further on, all vertices in the projected graph obtain the value of the former \texttt{"name"} property as label. The edge function in line 2 expresses that projected edges show only the original edge (type) labels while all edge properties are removed. In line 3, the projection operator is called on the input graph \texttt{db.G[0]} using the vertex and edge functions as arguments. The identifiers in the resulting new graph are temporary (e.g., \texttt{p0}) as projected graphs are typically reused in another operation. However, as stated above, it is also possible to persist temporary graphs.

\begin{algorithm}[h]
\caption{Projection Example}
\small
\begin{ttfamily}
\begin{algorithmic}[1]
\State vertexFunc = (Vertex v => 
\NoNumber \ \ new Vertex(v["name"], \{"from":v["city"]\})
\State edgeFunc = (Edge e => new Edge(e[:type], \{\}))
\State projGraph = db.G[0].\textbf{project}(vertexFunc,edgeFunc)
\end{algorithmic}  
\end{ttfamily}
\label{alg:project}
\end{algorithm}

\vspace{-2mm}
\paragraph*{\textbf{Summarization}}
The summarization operator determines a structural grouping of similar vertices and edges  to condense a graph and thus to help to uncover insights about patterns hidden in the graph \cite{zhang2010canal, zhao2011grapholap}. It can also be used as an optimization step to reduce the graph size with the intent to facilitate complex graph algorithms, e.g., multi-level graph partitioning \cite{karypis1998multilevel}.  The graph \textit{summarization} operator $\zeta_{g_v, g_e,\gamma_v,\gamma_e}:\mathcal{G} \rightarrow 	\mathcal{G}$ represents every vertex group  by a single vertex in the summarized graph; edges between vertices in the summary graph represent a group of edges between the vertex group members of the original graph.  Summarization is defined by specifying grouping keys $g_v$ and $g_e$ for vertices and edges, respectively, similarly as for \texttt{GROUP BY} in SQL.
These grouping keys are sets of property keys and may also include the type label $\tau$ (or \texttt{:type} in GrALa). Additionally, the vertex and edge aggregation functions $\gamma_v : \mathcal{V}^n \rightarrow \mathcal{V}$ and $\gamma_e : \mathcal{E}^n \rightarrow \mathcal{E}$ are used to compute aggregated property values for grouped vertices and edges, e.g.,  the average age of person groups or the number of group members, which can be stored at the summarized vertex or edge.

\begin{figure}[h]
\centering
\includegraphics[width=0.47\textwidth]{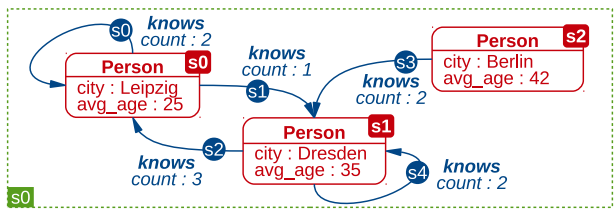}
\caption{Example summarization}
\label{fig:oex-summarization}
\end{figure}

Algorithm \ref{alg:summarize} shows an example application of our summarization operator using GrALa. The goal is to summarize persons in our graph of Fig. \ref{fig:epg-example} according to the city they live in and to calculate their average age. Furthermore, we want to group both the edges between users in different cities as well as edges between users that live in the same city. The result of the operator is shown in Figure \ref{fig:oex-summarization}. In line 1 we use the combine operator to form a single graph containing all persons and their relationships to each other; this will be the graph to summarize. In line 2 we define the vertex grouping keys. In this case, we want to group vertices by type label \texttt{:type} and property key \texttt{"city"}. Edges are only grouped by type label (line 3). Grouping keys and values are automatically added to the resulting summarized vertices and edges. In lines 4 and 5, we define the vertex and edge aggregation functions. Both receive the summarized entity (i.e., \texttt{vSum}, \texttt{eSum}) and the set of grouped entities (i.e., \texttt{vertices}, \texttt{edges}) as input. The vertex function applies the aggregate function \texttt{average(key)} on the set of grouped entities to compute the average age. The result is stored as a new property \texttt{avg\_age} at the summarized vertex. The edge function counts the grouped edges and adds the resulting value to the summarized edge. In line 6, the summarize operator is called using the predefined sets and functions as argument.

\begin{algorithm}[h]
\caption{Summarization Example}
\small
\begin{ttfamily}
\begin{algorithmic}[1]
\State personGraph = 
\NoNumber \ \ db.G[0].\textbf{combine}(db.G[1]).\textbf{combine}(db.G[2])
\State vertexGroupingKeys = \{:type,"city"\}
\State edgeGroupingKeys = \{:type\}
\State vertexAggFunc = (Vertex vSum, Set vertices =>
\NoNumber \ \ vSum["avg\_age"] = vertices.average("age"))
\State edgeAggFunc = (Edge eSum, Set edges =>
\NoNumber \ \ eSum["count"] = edges.count())
\State sumGraph = personGraph.\textbf{summarize}(
\NoNumber \ \ vertexGroupingKeys,vertexAggFunc,
\NoNumber \ \ edgeGroupingKeys,edgeAggFunc)
\end{algorithmic}  
\end{ttfamily}
\label{alg:summarize}
\end{algorithm}
\vspace{-5mm}

\vspace{-4mm}
\paragraph*{\textbf{Auxiliary Operators}}
\hspace{1mm}In addition to the fundamental graph and graph collection operators, advanced graph analytics often requires the use of application-specific graph mining algorithms. One application can be the extraction of subgraphs that cannot be achieved by pattern matching, e.g., the detection of communities in a social network \cite{fortunato2010communitydetection} or business transactions \cite{petermann2014biiig}. Further on, applications may require algorithms to detect frequent subgraphs \cite{jiang2013survey} or for statistical evaluations to select significant patterns. To support the plug-in of such external algorithms, we provide generic \textit{call} operators, which may have graphs and graph collections as input or output, formally $\eta_{a,P} : \mathcal{G} \cup \mathcal{G}^n \rightarrow \mathcal{G} \cup \mathcal{G}^n$. Depending on the output type, we distinguish between so-called  \texttt{callForGraph} (single graph result) and \texttt{callForCollection} operators. 
Algorithm \ref{alg:call} shows the use of \texttt{callForCollection} on a single input graph. The operator arguments are symbol $a$ to set the executed algorithm (e.g., \texttt{:CommunityDetection}) and a set of algorithm-specific parameters $P$. In the example, a \texttt{graphPropertyKey} needs to be supplied to determine, which graph property should store the computed community id. The resulting collection \texttt{communities} contains all logical graphs computed by the algorithm and can be used for subsequent analysis.

Furthermore, it is often necessary to execute an unary graph operator on more than one graph, for example to calculate an aggregated value for all graphs in a collection. Not only the previously introduced operators aggregation, projection and summarization, but all other operators with single graphs as in- and output (i.e., $o : \mathcal{G} \rightarrow \mathcal{G}$) can be executed on each element of a graph collection using the \textit{apply} operator $\lambda_o:\mathcal{G}^n \rightarrow \mathcal{G}^n$. For an input graph collection the specified operator is applied for every graph and the result is added to a new output graph collection. Algorithm \ref{alg:apply} demonstrates the apply function in combination with the aggregate operator. The latter is applied on all logical graphs in the database, represented by \texttt{db.G}. The result can be seen in our example graph in Figure \ref{fig:epg-example}, where each logical graph has an additional property for the vertex count.

\begin{algorithm}[h]
\caption{Call Example}
\small
\begin{ttfamily}
\begin{algorithmic}[1]
\State communities = graph.\textbf{callForCollection}(
\NoNumber \ \ :CommunityDetection,
\NoNumber \ \ \{"graphPropertyKey":"community"\})
\end{algorithmic}  
\end{ttfamily}
\label{alg:call}
\end{algorithm}
\vspace{-5mm}
\begin{algorithm}[h]
\caption{Apply Example}
\small
\begin{ttfamily}
\begin{algorithmic}[1]
\State db.G.\textbf{apply}(Graph graph => 
\NoNumber \ \ graph.aggregate(
\NoNumber \ \ \ \ "vertexCount",(Graph g => g.V.count()))
\end{algorithmic}  
\end{ttfamily}
\label{alg:apply}
\end{algorithm}
\vspace{-5mm}
\begin{algorithm}[h]
\caption{Reduce Example}
\small
\begin{ttfamily}
\begin{algorithmic}[1]
\State totalMerge = db.G.\textbf{reduce}(
\NoNumber \ \ (Graph g, Graph f => g.combine(f))
\end{algorithmic}  
\end{ttfamily}
\label{alg:reduce}
\end{algorithm}

Lastly, in order to apply a binary operator on a graph collection we adopt the \textit{reduce} operator often found in programming languages and also in parallel processing frameworks such as MapReduce. The operator takes a graph collection and a binary graph operator as input, formally $\rho_o : \mathcal{G}^n \rightarrow \mathcal{G}$. The binary operator $o : \mathcal{G}^2 \rightarrow \mathcal{G}$ is initially applied on the first pair of elements of the input collection which results in a new graph. This result graph and the next element from the input collection are then the new arguments for the binary operator and so on. In this way, the binary operator is applied on pairs of graphs until all elements of the input collection are processed and a final graph is computed. In Algorithm \ref{alg:reduce} we call the reduce operator parametrized with the combine operator on all logical graphs in the database in Figure \ref{fig:epg-example}. The final graph contains all persons of the three communities including their relationships to each other.

\section{Distributed Graph Store}
\label{sec:storage}

The distributed graph store is a fundamental element of the Gradoop framework. Its main purpose is to manage persistent EPGM databases by providing methods to read and write graphs, vertices and edges.\footnote{An API documentation is beyond the scope of this paper but will be provided in the documentation on our project website \url{www.gradoop.com}.} It further serves as data source and data sink for the operator implementations. 

The main requirement for a suitable implementation of the Gradoop graph store is supporting efficient access to very large EPGM databases with billions of vertices and edges including their respective properties. Graphs of that size can take up to multiple petabytes in space and thus require a distributed store that can handle such amounts of data. Furthermore, as already mentioned in section \ref{sec:architecture}, there should be different options to physically partition the graph data to ensure both load balancing as well as data locality with minimal communication overhead for graph processing. We also aim at supporting time-based graph analytics, so that the store should support data versioning of the graph structure as well as of properties. Finally, the store should provide 
fault tolerance against hardware failures and prevent data loss through data replication.

As there is currently no system to store and manage  graphs that apply to our data model, we chose Apache HBase \cite{hbase} as the technological platform for our distributed graph store. HBase is built on top of the Hadoop distributed file system (HDFS) and implements a distributed, persistent, multidimensional map. It can store large amounts of structured and semi-structured data across a shared nothing cluster and provides fast random reads and writes on that data to applications. Similar to relational databases, HBase organizes data inside tables that contain rows and columns. Unlike in the relational model, the table layout is not static as each row can have a very large number of different columns within column families. This leads to a very flexible storage layout optimized for sparse data and fits perfectly to the EPGM where each element can have various properties without following a global schema. 

The most basic unit in HBase is a cell which is identified by row key, column family, column and timestamp. Column families allow the grouping of columns based on their access characteristics and can be used to apply different storage features on them (e.g., compression or versioning). The timestamp enables data versioning at the cell level which is also supporting our requirements. HBase does not offer support for data types, instead, all values including the row key, column family, column and cell are represented by byte arrays leaving (de-)serialization of values to the application. To provide horizontal scalability in terms of data size and parallel data access, HBase partitions tables into so called regions 
and distributes them among cluster nodes. Built upon HDFS, a distributed, fault-tolerant file system, HBase also supports automatic failure handling through data replication. 

It can be seen, that Apache HBase already fulfills most of our stated requirements as it provides distributed management of large quantities of sparse data, data versioning and fault tolerance. 
A remaining challenge is to suitably map the EPGM to the data model provided by HBase. Furthermore, we should exploit the partitioning options of HBase for effective graph partitioning. In the following, we discuss our current implementation choices.

\paragraph*{\textbf{Graph Layout}}
Our current approach is straightforward: we represent the database graph as an adjacency list \cite{cormen2009algorithms} and store all vertices inside a single table (i.e., the \textit{vertex table}). Logical graphs are maintained in an additional \textit{graph table}. This approach gives us fast access to vertices including their properties and edges. It also gives us the possibility to quickly retrieve graphs including their corresponding vertex and edge sets.

\begin{figure*}[t]
\centering
\scriptsize
\begin{ttfamily}
\renewcommand{\arraystretch}{1.25}
\newcommand{\rowheadervertexstore}{\multicolumn{3}{|c|}{\textbf{meta}} & \multicolumn{4}{|c|}{\textbf{properties}} & \multicolumn{3}{|c|}{\textbf{out edges}} & \multicolumn{3}{|c|}{\textbf{in edges}} \\ \cline{2-14}}
\begin{tabular}{*{14}{|c}|}
\hline
\multirow{3}{*}{$0$-$0$} & \rowheadervertexstore
& type & graphs & idx & $name$ & $gender$ & \multicolumn{2}{|c|}{$city$} & \multicolumn{3}{|c|}{$\langle 2,0$-$1,0 \rangle$} & $\langle 2,0$-$1,0 \rangle$ & $\langle 3,0$-$9,0 \rangle$ & $\langle 4,0$-$9,2 \rangle$ \\ \cline{2-14}
& $0$ & $[0,2]$ & $1$ & $\langle 5,Alice \rangle$ & $\langle 5,f \rangle$ & \multicolumn{2}{|c|}{$\langle 5,Leipzig \rangle$} & \multicolumn{3}{|c|}{$[(since, \langle 0, 2014 \rangle)]$} & $[(since, \langle 0, 2014 \rangle)]$ & & \\ \hline

\multirow{3}{*}{$0$-$1$} & \rowheadervertexstore
& type & graphs & idx & $name$ & $gender$ & \multicolumn{2}{|c|}{$city$} & \multicolumn{3}{|c|}{$\langle 2,0$-$0,0 \rangle$} & \multicolumn{2}{|c|}{$\langle 2,0$-$0,0 \rangle$} & $\langle 3,0$-$9,1 \rangle$ \\ \cline{2-14}
& $0$ & $[0,2]$ & $1$ & $\langle 5,Bob \rangle$ & $\langle 5,m \rangle$ & \multicolumn{2}{|c|}{$\langle 5,Leipzig \rangle$} & \multicolumn{3}{|c|}{$[(since, \langle 0, 2014 \rangle)]$} & \multicolumn{2}{|c|}{$[(since, \langle 0, 2014 \rangle)]$} & \\ \hline

\multirow{3}{*}{$0$-$9$} & \rowheadervertexstore
& \multicolumn{2}{|c|}{type} & idx & \multicolumn{4}{|c|}{$title$} & $\langle 3,0$-$0,0 \rangle$ & $\langle 3,0$-$1,1 \rangle$ & $\langle 4,0$-$0,2 \rangle$ & \multicolumn{3}{|c|}{} \\ \cline{2-14}
& \multicolumn{2}{|c|}{$1$} & $3$ & \multicolumn{4}{|c|}{$\langle 5,Graph Databases \rangle$} & & & & \multicolumn{3}{|c|}{} \\ \hline
\end{tabular}
\end{ttfamily}
\caption{Vertex table containing the subgraph $\langle \{v_0, v_1, v_9\}, \{e_0, e_1, e_{15}, e_{17}, e_{18}\} \rangle$ of Figure \ref{fig:epg-example}.}
\label{fig:vertex-store}

\end{figure*}

\paragraph*{\textbf{Vertex table}}
The vertex space $\mathcal{V}$ and the edge space $\mathcal{E}$ of an EPGM database are stored in the vertex table. Each row contains all information regarding one vertex, i.e., vertex properties, incident edges including their properties and references to the graphs that contain the vertex. Typically, an operator implementation (e.g., in MapReduce) loads multiple rows from HBase and applies an algorithm on them. We store edges denormalized to give the operator implementations a holistic view on a vertex. By doing so, we avoid expensive join computation between the vertex table and a dedicated edge table during graph processing. Furthermore, given that HBase offers fast random access at the row level, our vertex store layout is advantageous for graph traversals as loading all incident edges of a vertex can be done in constant time.

Figure \ref{fig:vertex-store} shows a schematic representation of the vertex store omitting the time dimension. The graph contains three vertices resulting in three rows. Each row is identified by a row key which is the primary key within the table and composed of a \textit{partition id} and a \textit{vertex id}. Multiple rows may share the same partition id, but each vertex must have a unique vertex id which is either provided during data import or generated by the graph store. We will explain graph partitioning in more detail below.

Vertex data is further separated into four column families for two reasons. First, we assume that not all vertex data has the same access characteristics: while edges are frequently accessed in many graph algorithms, vertex properties on the other hand are only needed to evaluate predicate or aggregate functions. Second, HBase storage and tuning features are applied at the column family level, for example, compression requires similar column size characteristics to work more efficiently.

The column family \textit{meta} contains three columns at most. While the obligatory column \textit{type} stores the type label encoded by an id (e.g, \texttt{Person} is represented by 0), the second column \textit{graphs} stores the ids of graphs containing the vertex. The third column \textit{idx} stores an index which is used when creating outgoing edges (see below). If the vertex is not contained in any logical graph or has no outgoing edges, the particular columns do not exist and thus require no storage space.

The second column family \textit{properties} stores the vertex attributes. The number of grouped columns may differ significantly between rows as this depends solely on the vertex instance. The property key (e.g, $name$) is serialized as the column identifier while the property data type and the property value (e.g., $\langle 5, Alice \rangle$) are stored in the cell. As HBase solely handles byte arrays, the graph store adds support for all primitive data types (e.g., \texttt{String} is represented by 5). However, the property key does not enforce a specific data type for the associated value. Furthermore, data versioning is realized at the cell level and the number of versions is configurable in the Gradoop settings.

The remaining two column families store the incident edges of the vertex. Analogously to properties, the number of columns may vary significantly between rows. 
To enable efficient traversals in any direction, we currently store both outgoing and incoming edges per vertex. This leads to data redundancy as each edge has to be stored twice. However, the graph repository guarantees data consistency when updating edges. Each column in both column families serializes a single edge. The column stores an edge identifier, while the cell stores the edge properties. An edge identifier, e.g., $\langle 2,0$-$1,0 \rangle$, contains the edge type label (e.g., \texttt{knows} is represented by 2), the id of the opposite vertex (e.g., $0$-$1$) and an index which is unique at the start vertex. The opposite vertex identifier refers to the start- or end vertex of the edge depending on its direction. The edge index allows the definition of parallel edges. If an outgoing edge is created, the next available index is read from the \textit{idx} column and incremented afterwards. The graph store automatically adds the corresponding incoming edge at the target vertex using the same edge identifier with switched vertex ids. Edge properties are stored as a list of tuples, e.g., $[(since, \langle 0, 2014 \rangle)]$, where each tuple contains the property key, type and value. Consequently, reading a single edge property requires the deserialization of all edge properties. We decided to store edge properties differently from vertex properties as edges typically have significantly fewer properties than vertices. Similar to vertex properties, edge properties are versioned.

\begin{figure*}[t]
\centering
\scriptsize
\begin{ttfamily}
\renewcommand{\arraystretch}{1.25}
\newcommand{\rowheadergraphstore}{\multicolumn{2}{|c|}{\textbf{meta}} & \multicolumn{2}{|c|}{\textbf{properties}} & \multicolumn{4}{|c|}{\textbf{edges}} \\ \cline{2-9}}
\begin{tabular}{*{9}{|c}|}
\hline
\multirow{3}{*}{$0$} & \rowheadergraphstore
& type & vertices & $interest$ & $vertexCount$ & $0$-$0$ & $0$-$1$ & \multicolumn{2}{|c|}{$0$-$4$} \\ 	\cline{2-9}
& $5$ & $[0$-$0,0$-$1,0$-$4]$ & $\langle 5,Databases \rangle$ & $\langle 0,3 \rangle$ & $[\langle 2,0$-$1,0 \rangle]$ & $[\langle 2,0$-$0,0 \rangle]$ & \multicolumn{2}{|c|}{$[\langle 2,0$-$1,0 \rangle,\langle 2,0$-$2,1 \rangle]$} \\ 	\hline
\multirow{3}{*}{$1$} & \rowheadergraphstore
& type & vertices & $interest$ & $vertexCount$ & $0$-$2$ & $0$-$3$ & \multicolumn{2}{|c|}{$0$-$5$} \\ 	\cline{2-9}
& $5$ & $[0$-$2,0$-$3,0$-$5]$ & $\langle 5,Hadoop \rangle$ & $\langle 0,3 \rangle$ & $[\langle 2,0$-$3,0 \rangle]$ & $[\langle 2,0$-$2,0 \rangle]$ & \multicolumn{2}{|c|}{$[\langle 2,0$-$2,0 \rangle,\langle 2,0$-$3,1 \rangle]$} \\ 	\hline
\multirow{3}{*}{$2$} & \rowheadergraphstore
& type & vertices & $interest$ & $vertexCount$ & $0$-$0$ & $0$-$1$ & $0$-$2$ & $0$-$3$ \\ \cline{2-9}
& $5$ & $[0$-$0,0$-$1,0$-$2,0$-$3]$ & $\langle 5,Graphs \rangle$ & $\langle 0,4 \rangle$ & $[\langle 2,0$-$1,0 \rangle]$ & $[\langle 2,0$-$0,0 \rangle,\langle 2,0$-$2,1 \rangle]$ & $[\langle 2,0$-$1,0 \rangle,\langle 2,0$-$3,1 \rangle]$ & $[\langle 2,0$-$2,0 \rangle]$\\ \hline
\end{tabular}
\end{ttfamily}
\caption{Graph table containing the three logical graphs $\langle G_0, G_1, G_2 \rangle$ of Figure \ref{fig:epg-example}.}
\label{fig:graph-store}

\end{figure*}

\paragraph*{\textbf{Graph table}}
The set of logical graphs $\mathcal{G}$ of an EPGM database is stored in a second table, the graph table. Each row in that table contains all information regarding one graph, i.e., references to the vertices and edges it contains, a type label and properties. As illustrated in Figure \ref{fig:graph-store}, each row represents a single logical graph identified by a unique graph id and described by three column families. Similar to the vertex store, each row contains the column families \textit{meta} and \textit{properties}. While the former consists of the type label and a list of vertex identifiers contained in the graph, the latter stores graph properties in the same way as described for the vertex store. The third column family \textit{edges} stores all edges that are incident to the vertices contained in the logical graph. Each column stores a vertex identifier and the corresponding cell contains its outgoing edges belonging to the logical graph. This is necessary, as not all incident edges of a vertex may be contained in a logical graph. Furthermore, we can exploit the versioning features of HBase to load snapshots of logical graphs at a given time. While the column \textit{vertices} stores a versioned list of vertex identifiers, for each such identifier, we store a versioned list of incident edges, hence making the construction of structural snapshots possible.

\paragraph*{\textbf{Graph Partitioning}}
To achieve scalability of data volume and data access, HBase horizontally splits tables into so called regions and distributes those regions across the cluster. Each cluster node handles one or more regions depending on the available resources. Furthermore, rows inside a table are physically sorted by their row key, whereby each region contains a continuous range of rows between a defined start and end key. Region boundaries are either determined automatically or can be defined manually when creating a table. We apply the latter case to the vertex table when it is created and define partition boundaries upfront. For example, on a cluster with 10 nodes, an administrator may define 100 regions for the vertex table. Region boundaries are set by using the partition id, which is also used as the prefix in the row key.

Solely defining partition boundaries does not guarantee equal data distribution. To achieve that, the graph store supports partition strategies that assign a vertex to a region. At the moment, we support the well-known range and hash partitioning strategies, both requiring a continuous id space. The former assigns vertices to regions if their vertex id is in the partitions range, the latter assigns vertices to regions by applying a modulo function on the vertex id. Both strategies do not minimize the number of edges between different regions but achieve a balanced data distribution. We currently work on implementing more sensible strategies for improved locality of access.

\section{Use Case Evaluation}
\label{sec:usecases}
In this section, we present an initial evaluation of Gradoop for two analytical use cases, namely for social network analysis and business intelligence. 
We demonstrate the usefulness of the proposed data model and operators by showing that the non-trivial analysis tasks can be declared in relatively small GrALa scripts. 
To execute the equivalent workflows, we used initial implementations of the operators and generated the graph data by data generators. We first present the two analysis workflows and then discuss implementation and evaluation results for different graph sizes.  

\paragraph*{\textbf{Social Network Analysis}}
In our first scenario, an analyst is interested in communities of a social network. As a meaningful representation, she requires a summarized graph with one vertex per community, the number of users per community and the number of relationships between the different communities. Algorithm \ref{alg:sna} shows a GrALa workflow achieving such a summarized graph from a social network. The original social network graph \texttt{sng} is created using the LDBC-SNB Data Generator \cite{datagen,boncz2013ldbc} and shows different types of vertices and edges. However, communities should only group vertices of type \texttt{Person} and edges of type \texttt{knows}. So, in the first step, the relevant subgraph of \texttt{Person} vertices and \texttt{knows} edges is extracted in lines 1 to 4. In more detail, line 1 defines a \texttt{pattern} graph describing two vertices connected by one edge and line 2 the corresponding predicate. Then, in line 3 \texttt{pattern} and \texttt{predicate} are used to match all \texttt{knows} edges between \texttt{Person} vertices. The result is \texttt{friendships}, a collection of 1-edge graphs, which is subsequently combined to a single graph \texttt{knowsGraph} utilizing the reduce operator. In line 5, an external algorithm \texttt{:LabelPropagation} \cite{raghavan2007labelpropagation} is executed to detect communities. We use the \texttt{callForGraph} operator, as we need a single graph as output, where each vertex has a \texttt{"community"} property. Finally, the graph is summarized in line 6. At that, vertices are grouped by \texttt{"community"}. As the set of edge grouping keys is empty, edges are only grouped by their incident vertices. Both, vertices and edges in the summarized graph provide a \texttt{"count"} property showing the number of original vertices or edges. 

\begin{algorithm}[t]
\caption{Summarized Communities}
\small
\textbf{Input:} Social Network Graph \texttt{sng} \\
\textbf{Output:} Summarized graph \texttt{summarizedCommunities}; each vertex represents a community and edges represent aggregated links between communities

\begin{ttfamily}
\begin{algorithmic}[1]
\State pattern = new Graph("(a)-c->(b)")
\State predicate = (Graph g => 
\NoNumber \ \ g.V[\$a][:type] == "Person" \&\&
\NoNumber \ \ g.E[\$c][:type] == "knows" \&\& 
\NoNumber \ \ g.V[\$b][:type] == "Person") 
\State friendships = sng.\textbf{match}(pattern,predicate)
\State knowsGraph = friendships.\textbf{reduce}(
\NoNumber \ \ Graph g, Graph f => g.\textbf{combine}(f))
\State knowsGraph = knowsGraph.\textbf{callForGraph}(
\NoNumber \ \ :LabelPropagation,\{"propertyKey":"community"\})
\State summarizedCommunities = knowsGraph.\textbf{summarize}(
\NoNumber \ \{"community"\}, (Vertex vSum, Set vertices) =>
\NoNumber \ \ \ \ vSum["count"] = vertices.count()),
\NoNumber \ \{\},(Edge eSum,Set edges)=> 
\NoNumber \ \ \ \ eSum["count"]=edges.count())
\end{algorithmic}
\end{ttfamily}
\label{alg:sna}
\end{algorithm}
\ 

\paragraph*{\textbf{Business Intelligence}}
In our second scenario, an analyst is interested in common data objects, such as employees, customers or products, occuring in all high turnover business transaction graphs (business process executions) \cite{petermann2014biiig}. Algorithm \ref{alg:btgs} shows a corresponding GrALa workflow. The initial integrated instance graph \texttt{iig} contains all domain objects and relationships determined by the FoodBroker data generator \cite{petermann2014foodbroker}. In line 1, a domain specific algorithm is executed to extract a collection of business transaction graphs (\texttt{btgs}) using the algorithm from \cite{petermann2014biiig}. Line 2 defines an advanced predicate to select  the revenue-relevant graphs containing at least one vertex with type label \texttt{Invoice}. Line 3 defines an advanced aggregation function to calculate the actual revenue per graph. In line 4, multiple operations are chained. First, only the graph meeting the \texttt{predicate} are selected and second, the actual revenue is aggregated and written to the new graph property \texttt{"revenue"} for all remaining graphs. In line 5, the graphs with top 100 revenue aggregates are selected using our sort and top operators. Finally, the overlap of all subgraphs is determined by applying our reduce operator in line 6.

\begin{algorithm}[t]
\caption{Top Revenue Business Cases}
\small
\textbf{Input:} integrated instance graph \texttt{iig} \\
\textbf{Output:} Common subgraph of top 100 revenue business transaction graphs \texttt{topBtgOverlap}
\begin{ttfamily}
\begin{algorithmic}[1]
\State btgs = iig.\textbf{callForCollection}(
\NoNumber \ \ :BusinessTransactionGraphs,\{\})
\State predicate = (Graph g =>
\NoNumber \ \ g.V.\textbf{select}(Vertex v =>
\NoNumber \ \ \ \ v[:type] == "SalesInvoice").count() > 0)
\State aggRevenue = (Graph g => 
\NoNumber \ \ g.V.values("revenue").sum())
\State invBtgs = btgs.\textbf{select}(predicate)
\NoNumber \ \ .\textbf{apply}(Graph g => 
\NoNumber \ \ \ \ g.\textbf{aggregate}("revenue",aggRevenue))
\State topRevBtgs = invBtgs
\NoNumber \ \ .\textbf{sortBy}("revenue",:desc).\textbf{top}(100)
\State topRevBtgOverlap = invBtgs.\textbf{reduce}(
\NoNumber \ \ Graph g, Graph h => g.\textbf{overlap}(h))
\end{algorithmic}  
\end{ttfamily}
\label{alg:btgs}
\end{algorithm}

\pagebreak

\paragraph*{\textbf{Implementation and evaluation}}

For the initial evaluation and proof-of-concept we implemented the required operators in Giraph and MapReduce and use the HBase graph store as data source and sink between operator executions. Before the workflow runs, the generated data set is loaded into HBase using its MapReduce bulk import. Vertices are assigned to regions using the range partitioning strategy which leads to a balanced distribution. The matching and combination steps are realized by loading the relevant subgraph from the graph store. We further implemented the Label Propagation algorithm for the first use case and the extraction of business transaction graphs for the second use case in Giraph. Selection, aggregation and summarization have been implemented in MapReduce. As HBase does not natively support secondary indexes, we implemented the sort operator by building a secondary index during workflow execution with MapReduce. The top and overlap operators access the graph store directly to select relevant graphs and their elements. Both operators are implemented as regular, non-distributed  Java applications. Generally, the integration of different frameworks, i.e., HBase, MapReduce and Giraph, could be easily done as they belong to the same ecosystem and thus share libraries and internal approaches like data serialization.

Table \ref{tab:eval} shows the results of our initial evaluation on a small cluster of five nodes, each equipped with an Intel Xeon CPU E5-2430, 48 GB RAM and local disk storage. In both use cases, the data generators were executed with their default parameters. We adjusted the scale factors (SF) to generate graphs of different sizes. The resulting sizes are shown in the table as well as the time to load them from HDFS into the graph store. We observe that for both use cases the loading times scale linearly with the graph size. The execution times for the analytical workflows in the rightmost column also show a linear and thus scalable behavior w.r.t. the graph sizes. 
Nevertheless, we observed that the time to read and write data from HBase needs to be further optimized, e.g., for Giraph-based operators where about 50\%  of the processing time was for loading and distributing the graph. To improve this, we need to replace Giraphs own partitioning strategies to avoid data transfers across the cluster while loading the graph.

%

\begin{table}[h]
\centering
\footnotesize
\begin{tabular}{*{6}{r}}
\toprule
\multicolumn{1}{c}{Datagen}					&
\multicolumn{1}{c}{SF}						&
\multicolumn{1}{c}{$|V|$}					&
\multicolumn{1}{c}{$|E|$} 					&
\multicolumn{1}{c}{Import [s]}				& 
\multicolumn{1}{c}{\textbf{Workflow [s]}} 	\tabularnewline 
\midrule
LDBC-SNB 	& 1		&	3.6M 	& 21.7M & 79	& 218 \tabularnewline 
\cmidrule{1-6}
LDBC-SNB 	& 10	&	34M 	& 217M 	& 828 	& 1984 \tabularnewline 
\midrule
FoodBroker 	& 100	&	7M 		& 70M 	& 259 	& 234 \tabularnewline 
\cmidrule{1-6}
FoodBroker 	& 1000	&	70M 	& 700M 	& 2020 	& 1754 \tabularnewline
\bottomrule
\end{tabular}
\caption{Statistics for both use cases.}
\label{tab:eval}
\end{table}

\section{Related Work}
\label{sec:related}

We discuss related work on graph data models as well as on systems and approaches for graph data management and analysis. We also deal with more specific recent work on graph analytics related to our operators.

A large variety of graph data models has been proposed in the last three decades \cite{angles2012comparison,angles2008survey, dries2012biql,he2008graphs}, but only two found considerable attention in graph data management and processing: the \textit{resource description framework (RDF)} \cite{klyne2006rdf} and the \textit{property graph model (PGM)} \cite{rodriguez2010propertygraph, rodriguez2010traversalpattern}. In contrast to the PGM, RDF has some support for multiple graphs by the notion of n-quads \cite{cyganiak2008n}; its  standardized query language SPARQL \cite{sparql} also allows queries on  multiple graphs. However, the RDF data representation by triples is very fine-grained and there is no uniform way to represent richer concepts of the PGM in RDF \cite{souripriya2014oracle} so that the distinction of relationships , e.g., \texttt{(vertex1,edge1,vertex2)}, type lables, e.g., \texttt{(edge1,type,knows)}, and properties, e.g., \texttt{(vertex1,name,\\Alice)} has to be done at the application level. In consequence, expressing queries involving structural and value-based predicates requires non-standard extensions \cite{sakr2012gsparql}.

Gradoop provides a persistent graph store and an API to access its elements
similar to graph database systems, e.g., Neo4j \cite{neo4j} and Sparksee \cite{sparksee,martinez2011dex}.  In contrast to most graph database systems, the Gradoop store is built on a distributed storage solution and can partition the graph data across a cluster. A notable exception is Titan \cite{titan}, a commercial distributed graph database supporting different storage systems, e.g., Apache Cassandra or HBase. Unlike Gradoop, Titan focuses on transactional graph processing and the storage layout is built for the PGM. Approaches to store and process RDF data in Hadoop are surveyed in \cite{kaoudi2015rdf}.

MapReduce is heavily used for the parallel  analysis  of voluminous data \cite{shim2012mapreduce} and has also been applied for iterative algorithms \cite{dittrich2013efficient,lin2010graphmapreduce}. 
GLog \cite{gao2014glog} is a promising graph analysis system that extends datalog with additional rules for graph querying. GLog queries are translated to a series of optimized MapReduce jobs. The underlying data model is a so called relational-graph table (RG table)  storing vertices by nested attributes. Unlike Gradoop, GLog stores RG tables as HDFS files and does not address graph partitioning and data versioning in general. It also lacks more complex graph analytics on multiple graphs and, as noted by the authors, suffers from general limitations of MapReduce for iterative algorithms that can be reduced by graph processing systems. 

Parallel graph processing systems, such as Pregel \cite{malewicz2010pregel}, Giraph \cite{giraph}, GraphLab\cite{low2012graphlab} and the recent Pregelix \cite{yingyi2015pregelix}, focus on the efficient, distributed execution of iterative algorithms on big graphs. The provided data models are generic and algorithms need to be implemented by user-defined functions. Gradoop can use these systems by mapping its high-level, declarative operators to their respective generic data-models and user-defined functions.
Parallel processing systems such as Spark \cite{zaharia2010spark} and Flink \cite{flink} (formerly known as Stratosphere \cite{alexandrov2014stratosphere}) support analysis workflows with high-level graph operators, similarly as in Gradoop. However, their graph data models and graph operators are limited to single graphs. For example, Spark GraphX \cite{xin2013graphx} and Flink Gelly support filter operations on vertex and edge sets to extract a subgraph from a single graph, but no further analytical operations on graph collections. Specific graph algorithms, e.g., connected components or page rank, are implemented as dedicated operators in GraphX and Flink. The mentioned graph processing systems focus on in-memory processing and do not provide a persistent, distributed graph store like Gradoop. Still, we see Spark and Flink as powerful and supportive platforms for our approach and we plan to make use of them.

Shared memory cluster systems such as Trinity \cite{shao2013trinity} address both online query processing and offline analytics on distributed graphs. Trinity demands a strict schema definition and has no support for graph collections. The system offers an API to access graph elements and leaves the implementation of analytical operators to the user. Furthermore, Trinity is an in-memory system with no support for persistent graph data management.

There are also graph analytics tools built on top of relational database systems thereby utilizing their proven performance techniques for query processing.  
For example, Vertexica \cite{jindal2014vertexica} offers a vertex-centric query interface for analysts to express graph queries as user-defined functions which are executed by a SQL engine. In contrast to Gradoop, relational graph stores may be less suited to support schema-flexible graph models such as the EPGM and are not as well integrated in the Hadoop ecosystem to utilize its potential for parallel graph mining.

Many publications propose specific implementations related to some of our operators, although we can only discuss some of them here.  Similar to our summarization operator, Rudolf et al. \cite{rudolf2013synopsis} describe a visual approach to declaratively define graph summaries. Liu et al describe distributed algorithms for graph summarization \cite{liu2014distributed}. OLAP-like graph analysis using multiple summaries of a graph is proposed among others in \cite{colazzo14rdf,zhao2011grapholap}. Some approaches support heterogeneous graphs \cite{yin2012hmgraph}, grouping by edge attributes \cite{wang2014pagrol} and the generation of a predefined number of vertex groups \cite{tian2008efficient, zhang2010canal}. Summaries are not only useful as a simplified representation of a graph, but also to optimize queries \cite{lefevre2010grass,riondato14summarization}. 
While pattern matching queries \cite{angles2008survey,angles2012comparison} are a typical part of existing graph data models, recent work is focussing on pattern queries in graph collections \cite{bleco2014graph}, distributed graphs \cite{fan14simulation} and by graph similarity \cite{zhao2013efficient}.

\vspace{-2mm}
\section{Conclusions}
\label{sec:conclusion}
We presented an overview of Gradoop, an end-to-end approach for Hadoop-based management and analysis of graph data.
Its underlying extended property graph data model (EPGM) builds upon the proven property graph model but extends its with
support for collections of graphs which we expect to be a major asset in many analytical applications. 
The proposed set of operators provides basic analysis and aggregation capabilities, graph summarization and collection processing.
All operators as well as the invocation of  graph mining algorithms  are usable within the GrALa  language to specify analysis scripts.
The Gradoop store is realized based on HBase and supports scalability to very large graphs, graph versioning, partitioned storage and fault tolerance. An initial evaluation shows the flexibility of the proposed operators to define analytical workflows in different domains as well as the scalability of the parallel data import and workflow execution for different graph sizes. 

Gradoop is still in its initial phase so that many parts of the system need to be completed and optimized, in particular operator implementations, the upper layers of the architecture (Figure \ref{fig:architecture}) and data integration workflows. For the efficient implementation of the workflow execution layer and operators we will evaluate and possibly utilize available  Flink and GraphX functionality. We will also address component-specific  research questions such as customizable graph partitioning  at different storage layers (HBase, in-memory) and the optimization of specific operators and entire workflows. Gradoop will be open-source and made available under www.gradoop.com. 


 
\section{Acknowledgments}

This work is partially funded by the German Federal Ministry of Education and Research under  project ScaDS Dresden/Leipzig (BMBF 01IS14014B).

\footnotesize
\bibliographystyle{abbrv}
\bibliography{main}

\end{document}